# A Change Language for Ontologies and Knowledge Graphs

Harshad Hegde[1], Jennifer Vendetti[2], Damien Goutte-Gattat[3], J Harry Caufield[1], John B Graybeal[2], Nomi L Harris[1], Naouel Karam[4], Christian Kindermann[2], Nicolas Matentzoglu[5], James A Overton[6], Mark A Musen[2], Christopher J Mungall[1,*]

[1]Lawrence Berkeley National Laboratory, Berkeley, CA, 94720, USA
[2]Stanford University, Palo Alto, CA 94304, USA
[3]University of Cambridge, Cambridge CB2 3DY, UK
[4]Institute for Applied Informatics (InfAI), Leipzig University, 04109 Leipzig, Germany
[5]Semanticly, Athens, Greece
[6]Knocean Inc., Toronto, Ontario, Canada

* Corresponding author, cjmungall@lbl.gov

## Abstract

Ontologies and knowledge graphs (KGs) are general-purpose computable representations of some domain, such as human anatomy, and are frequently a crucial part of modern information systems. Most of these structures change over time, incorporating new knowledge or information that was previously missing. Managing these changes is a challenge, both in terms of communicating changes to users, and providing mechanisms to make it easier for multiple stakeholders to contribute.

To fill that need, we have created **KGCL, the Knowledge Graph Change Language**, a standard data model for describing changes to KGs and ontologies at a high level, and an accompanying human-readable controlled natural language. This language serves two purposes: a curator can use it to request desired changes, and it can also be used to describe changes that have already happened, corresponding to the concepts of "apply patch" and "diff" commonly used for managing changes in text documents and computer programs. Another key feature of KGCL is that descriptions are at a high enough level to be useful and understood by a variety of stakeholders ─ for example, ontology edits can be specified by commands like "add synonym 'arm' to 'forelimb'" or "move 'Parkinson disease' under 'neurodegenerative disease'".

We have also built a suite of tools for managing ontology changes. These include an automated agent that integrates with and monitors GitHub ontology repositories and applies any requested changes, and a new component in the BioPortal ontology resource that allows users to make change requests directly from within the BioPortal user interface.

Overall, the KGCL data model, its controlled natural language, and associated tooling allow for easier management and processing of changes associated with the development of ontologies and KGs.

# Introduction

Ontologies are structures that encode concepts and entities in a particular domain in a way that facilitates data standardization as well as a wide variety of inferential tasks. They are crucial to many modern information systems. Ontologies such as the Gene Ontology (GO) are used daily to interpret high-throughput experimental data. Anatomical and cell type ontologies such as the Cell Ontology (1) and Uberon (2) are crucial for projects like HuBMAP (3) and the Human Cell Atlas (4) that aim to provide a molecular map of the bodies of humans and other organisms. In the clinical realm, ontologies such as the Human Phenotype Ontology (5) are being used to standardize the representation of phenotypes in patients, and for supporting applications such as phenotype-based variant prioritization. Ontologies can be distributed and accessed via portals such as BioPortal (6), OntoBee (7), and the Ontology Lookup Service (8).

Ontologies often have a graph-like structure, and are frequently incorporated into knowledge graphs (KGs), which augment the textbook knowledge in an ontology with additional evidence-backed knowledge about individual entities. Examples of KGs in the biosciences include Hetionet (9), PheKnowLator (10), and KG-COVID-19 (11). Ontologies and KGs are similar structures, with different emphases. Ontologies may emphasize expressive logical structures, and KGs typically emphasize simpler interconnections.

While ontologies and KGs vary tremendously in their applications, structure, and formalisms, a common underlying element is *change*. They are not static. It is rare that a domain is ever represented with perfect accuracy and completion on a first pass – instead there is a constant process of refinement, as part of a complete lifecycle, involving many stakeholders. **Figure 1**, which shows the number of changes in GO terms between releases, illustrates this continual process. GO was first released over 20 years ago, yet its rate of change remains high and even increases, as the available pool of relevant knowledge continues to grow.

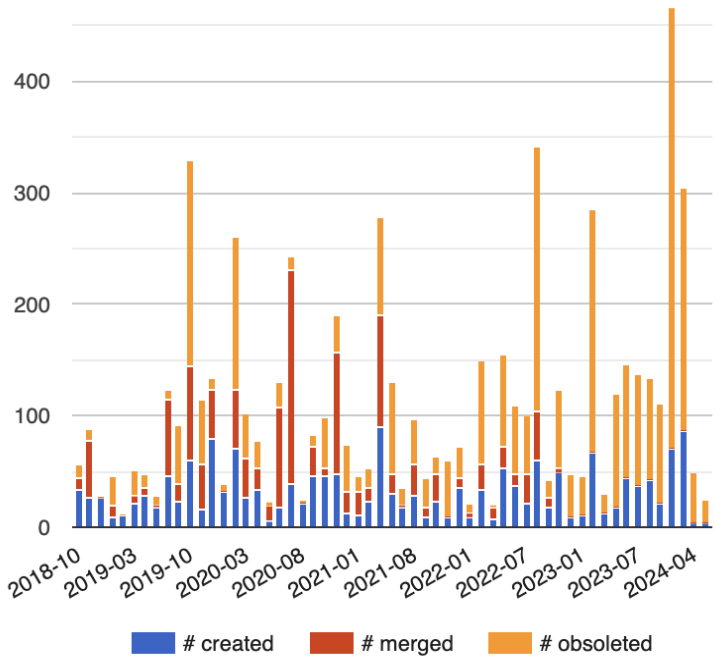

**Figure 1:** Changes in GO terms (number created, merged, or obsoleted) between major releases between late 2018 and early 2024. Biomedical ontologies such as GO change frequently in response to new knowledge.

Changes to ontologies generally fall into several categories: adding a new element, obsoleting an existing element, merging two elements, and adding/deleting/modifying information associated with an element (such as definitions, synonyms or taxon restrictions). Changes can happen in response to new knowledge, new terminology, improved modeling of knowledge, or simply to correct previous errors. Changes can be instigated by requests from the community, but they are typically applied by expert curators and ontology editors.

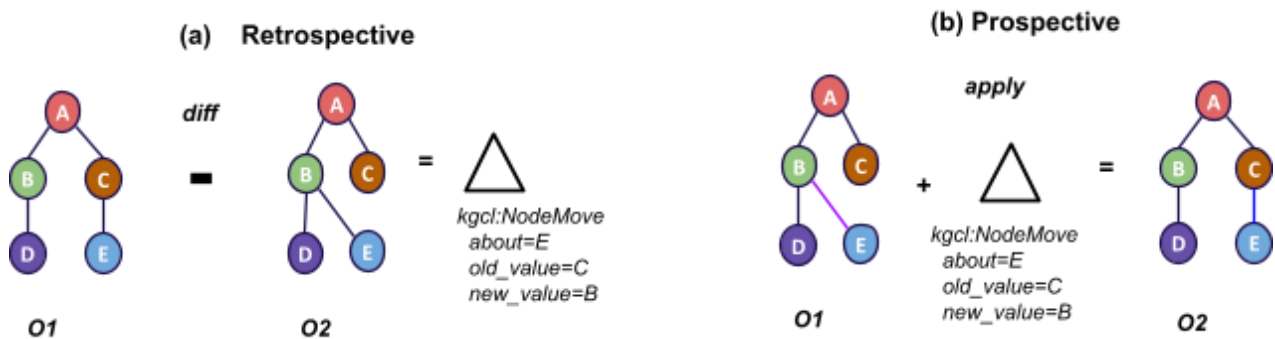

**Figure 2**: Logic of ontology changes. (a) Retrospective: comparing two ontologies O1 and O2 via a *diff* operation yields a Change object representing the most parsimonious set of changes to go from O1 to O2; in this case, the change is moving the node E from under node C to under node B. (b) Prospective: *applying* a Change object to O1 yields O2. The *diff* operation can be used to describe retrospective changes that have happened to an ontology over time. The *apply* operation can be used to describe intended changes to an ontology prospectively.

Despite the constant factor of change, there is surprisingly no single agreed-upon way to *represent or communicate* ontology changes. This is a major technical and communication obstacle for KGs and ontologies. This obstacle presents in two ways (**see Figure 2**). The first is how changes that have been made *retrospectively* are communicated to stakeholders (i.e. "diffs"). The second way is how contributors communicate *prospective* desired changes to the KG (i.e. "patches"). A typical workflow is for a domain expert or curator to ask for a change in natural language, sometimes in the form of an issue/ticket in the GitHub project for an ontology. A specialized ontology editor then translates this request into a sequence of actions in an ontology development environment, such as Protege (12) (13) or WebProtege (14). This is a repetitive and time-consuming task with many inefficiencies, and it relies on the availability of ontology editors to process these change requests. Another challenge in ontology editing is that different ontologies implement different workflows. Sometimes these are documented, and sometimes one ontology will partially or fully adopt procedures from another ontology, but overall there is a lack of commonality which makes it harder to automate and for curators to transfer practice from one ontology to another. Standardizing these processes can reduce the number of mistakes and increase efficiency.

## Community need for a change language and associated tools

In order to better understand what curators and ontology developers need from a change language, we hosted a virtual workshop in 2023 on change languages in ontologies. In order to scope the workshop we focused on the kinds of biological ontologies found in BioPortal. Before and during the workshop we surveyed the community on a range of topics, including what tools they used to generate ontology diffs, how satisfied they were with those tools, and how they made changes to ontologies. We also demonstrated and solicited feedback on a proposed standard for representing ontology changes (which became KGCL).

In our survey of ontology users, when asked how important it was to them to stay informed about changes to ontologies they use, 82% rated it extremely or very important. The two tools that participants mentioned for generating diffs were the ROBOT ontology tool (15), and the Bubastis ontology diff tool (16). ROBOT offers an ontology diff operation among its many different ontology processing operations, and can operate over ontologies in OBO format or any OWL syntax. Bubastis is a dedicated ontology diffing tool that has been integrated into BioPortal and other OntoPortal endpoints. It is automatically executed for each new ontology release, allowing users to easily download and view changes. Both operate by performing setwise diff of OWL axioms, which gives a precise computable representation of changes, but at a lower level than how many ontology developers conceive of changes. For example, the NodeMode operation in **Figure 2** would be represented as two changes, a deletion and an insertion. In our surveys users reported they would be more satisfied with a higher level representation and presentation of changes.

In addition to ROBOT and Bubastis, other ontology diffing frameworks include the QuickGO Change Log (17), GOtrack (18), and the COnto-diff framework (19). Although these were not mentioned by survey participants in the list of tools they use, these three frameworks provide complementary and powerful ways of viewing and understanding changes in ontologies. Both the QuickGO Change Log and GOtrack are specific to the Gene Ontology (GO). The QuickGO Change Log is integrated into the QuickGO ontology browser, and allows users to see changes to the GO in the context of other information and annotations about that term. GOtrack allows users to analyze the impact of changes on the GO. COnto-diff provides ontology diffing for any OWL ontology, and pioneered the use of a taxonomy or classification of change types (see section "Aligning with related work").

The change language workshop also covered the application of changes to ontologies. When we surveyed the community of people who use and/or build ontologies and asked how they request changes in an ontology, almost all of them were following the workflow just described, though some were technically skilled enough to make Pull Requests in the ontology repository to accomplish the desired changes. Only 17% said they were very or extremely satisfied with the turnaround time for the changes to happen.

Based on feedback from the workshop participants, there was a clear need for a standardized high-level way of representing changes in ontologies, as well as improved methods for being able to describe changes to automated agents that could apply these changes seamlessly to existing ontologies, and that similar methods could be applied to knowledge bases and KGs more generally.

## KGCL provides a standard for describing ontology changes

To address the need for a standardized way to express changes in ontologies, we created KGCL, the Knowledge Graph Change Language. KGCL can represent common ontology editing operations (such as modifying a label or a definition, obsoleting a term, moving a term under another parent term, etc.) using a Controlled Natural Language (CNL). This CNL is designed to be as close to natural language (US English) as possible, yet to be unambiguously parseable by machines. As an example, the KGCL command to change the name of the ontology term with the ID ENVO:01000575 from 'wax' to 'oil' is "rename ENVO:01000575 from 'wax' to 'oil'".

KGCL consists of three primary components (see **Figure 3**):
1. A schema and taxonomy (classification) of change types
2. A Controlled Natural Language (CNL) for specifying ontology changes
3. Multiple serialization formats such as JSON, YAML, RDF, as well as tabular formats for use in spreadsheets.

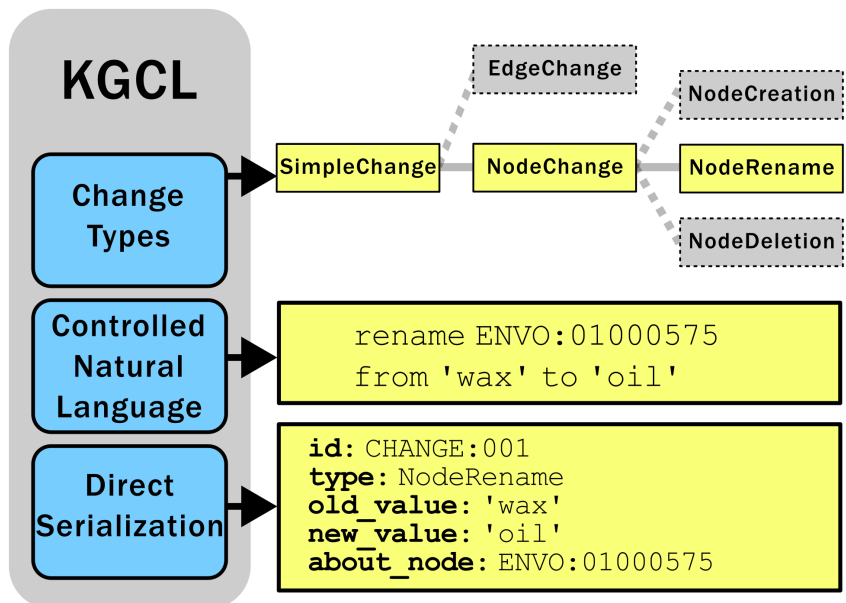

**Figure 3: Overview of the three components of KGCL.** (1) A classification of change types (here showing that NodeRename is a subtype of NodeChange, and a sibling of NodeCreation and NodeDeletion). (2) A CNL for expressing changes in a simple human-readable yet computable syntax. (3) A data model that can be directly serialized in different syntaxes (here showing a NodeRename instance serialized as YAML).

## A classification of types of ontology changes

KGCL organizes different kinds of changes into a classification hierarchy, such that all changes that affect terms (nodes) are in one branch, and all changes that affect relationships (edges) are in another branch – see **Table 1**.

**Table 1:** Types of changes supported by KGCL, grouped into Node Changes and Edge Changes, with example KGCL command for each type, using the Uberon anatomy ontology.

| Change type | Description | Example Command |
|---|---|---|
| **NodeChange** | | |
| NodeRename | A node change where the name (aka rdfs:label) of the node changes | rename UBERON:0002398 from 'hand' to 'manus' |
| NodeObsoletion | Deprecates usage of the node, but does not delete it | obsolete 'trachea' (alternatively: obsolete 'UBERON:0003126') |
| NodeDeletion | Deletes a node from the graph | delete node 'heart' |
| ClassCreation | A node creation where the node is a class (as opposed to a relation) | create 'digestive system' |
| Synonym Replacement | A node synonym change where the text of a synonym is changed | replace synonym 'intestine' with 'gut' for 'alimentary canal' |
| NewTextDefinition | A node change where a de-novo text definition is created | add definition 'A muscular organ that pumps blood through the body' to 'heart' |
| RemoveTextDefinition | A node change where a text definition is deleted | remove definition for 'liver' |
| NodeTextDefinitionChange | A node change where the text definition is changed | change definition of 'kidney' to 'An organ that filters blood to produce urine' |
| NewSynonym | A node synonym change where a de-novo synonym is created | create exact synonym 'thigh bone' for 'femur' |
| RemoveSynonym | A node synonym change where a synonym is deleted | remove synonym 'arm bone' for 'humerus' |
| **EdgeChange** | | |
| EdgeCreation | An edge change in which a de-novo edge is created | create edge 'hepatocyte' part_of 'liver' |
| EdgeDeletion | An edge change in which an edge is removed | delete edge 'hepatocyte' part_of 'lung' |
| NodeMove | A combination of deleting a parent edge and adding a parent edge | - |
| PredicateChange | An edge change where the predicate (relationship type) is modified | change relationship between 'stomach' and 'digestive system' from 'is_a' to 'part_of' |

## Change Data Model

We expressed the above hierarchy in a semantic data model. The data model describes the attributes of each change type. Some attributes, such as the `id` attribute (a unique identifier for tracking each change), are shared across all change types. Other attributes are specific to individual types of change.

For example, as shown in **Figure 3**, the NodeRename class has an attribute 'about_node' (shared by all NodeChange objects, describing the node or term to be acted on), as well as "old_value" and "new_value", describing the name/label to be changed and the replacement name/label.

We use LinkML (20) to express the KGCL Data Model. This allows for flexible modeling features such as class hierarchies, as well as providing a semantic representation of the model. LinkML also allows the data model to be expressed using other frameworks, such as OWL or

JSON-Schema. In addition, we used the LinkML tool chain to derive the KGCL website (https://w3id.org/kgcl) from the data model.

## Serialization formats and Controlled Natural Language (CNL) expression

KGCL can be serialized and deserialized using different syntaxes. The canonical syntax for KGCL is the KGCL Controlled Natural Language (CNL), which is intended to be easily read and written by humans, but is also parseable by machines. The KGCL CNL is specified by a grammar using the Lark formalism (21). KGCL can be serialized as JSON, YAML, or RDF, if there is no need for human readability. A tabular form is also available for use in spreadsheets, but this is less expressive than the other forms.

# A tool suite for working with KGCL

We have built tools aimed at ontology developers, curators, and software developers to help with common tasks related to change management. These tools include an automated agent (Ontobot) that waits for requests from curators on GitHub issue trackers, and then enacts these changes on an ontology; a widget for the BioPortal ontology portal that allows users to make change requests in the BioPortal user interface; and software libraries and command line tools in Java and Python that can be used by advanced users.

## Ontobot: an automated agent for applying curator change requests

Many ontologies are managed in GitHub (22), with GitHub issues used to manage change requests from users, and GitHub Pull Requests (PRs) to suggest these changes. This is currently a manual and time-intensive process, in which an ontology editor will read through ontology issues, carry out the requested changes using an ontology development tool such as Protégé, and then make a PR, which is later reviewed and merged. We created an automated agent called Ontobot that simplifies this workflow by automating the time-consuming intermediate steps in this workflow.

Ontobot is integrated with GitHub using the GitHub Actions mechanism (23). The maintainers of an ontology repository can easily deploy Ontobot, after which it will monitor the GitHub repo, where it watches for issues with a specific text string: "*Hey ontobot! apply:*", followed by a bulleted list of ontology change requests, written in the KGCL CNL syntax. The agent will then carry out the request, generating a GitHub PR that will make the requested change(s) in the ontology source file. The PR can be quickly reviewed and merged by the maintainers of the ontology. **Figure 4** shows an example of this process in which a user requests the addition of a synonym to a term in the Mondo Disease Ontology (MONDO) and Ontobot creates a PR to make the change in the ontology file, which needs to be approved by a Mondo curator.

Step 1: User creates a GitHub issue requesting an ontology change (using the KGCL CNL)

Step 2: Ontobot reads the issue and creates a Pull Request to make the change(s)

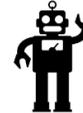

Step 3: The diff is examined by an ontology developer

Step 4: The proposed change is merged into the ontology

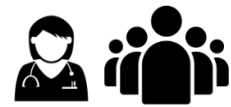

**Figure 4: An example of a user-initiated change request handled by Ontobot.** Ontobot can be invoked by a user via a GitHub issue to make a change to an ontology (in this case, adding a synonym to a MONDO term).
**Step 1**: The user opens a GitHub issue in the ontology repository requesting an ontology change. The issue includes the special instruction "## Hey ontobot! apply:" followed by commands in KGCL CNL syntax describing the desired change(s) (here, adding an exact synonym to a term).
**Step 2**: The Ontobot change agent, which watches the issue tracker for the "Hey ontobot" instruction, sees this issue and responds to it.
**Step 3**: Ontobot creates a Pull Request (PR) that will execute the requested change to the ontology. Curators are assigned to review the PR; it cannot be merged until at least one curator approves it.
**Step 4:** Once the curator approves the proposed change, it is merged into the ontology and incorporated into the next release, where it is available for use by all.

## User-friendly suggestion of ontology changes via BioPortal

The BioPortal ontology portal provides access to a wide selection of biomedical ontologies, and tools for working with them. A new BioPortal widget, still in development, provides an easy way for biocurators to suggest changes to an ontology they work on. Under the hood, this widget uses KGCL CNL to express the desired ontology changes and invokes Ontobot to create the corresponding pull request in GitHub. As we showed in the last section, biocurators can go directly to GitHub and open an issue to call Ontobot; the BioPortal widget provides an even simpler and more user-friendly interface that does not require any knowledge of KGCL.

When BioPortal users browse ontology classes, the new widget provides easy access to forms in the user interface for entering information about proposed changes (see **Figure 5**). Each form presented to users has the necessary fields to collect data that is specific to the various change types – e.g., for the addition of a synonym, a dropdown field allows the user to specify the type of synonym such as exact, narrow, broad, or related.

**Step 1:** In BioPortal, the user looks at an ontology term. They want to suggest a new synonym, so they click the "+" button next to "Synonyms".

**Step 2:** BioPortal pops up a change form so the user can make their change request

**Step 3:** BioPortal submits a GitHub issue on the user's behalf

**Figure 5: Requesting an ontology change in BioPortal.** In this example, the user opens the Mondo term "neurodevelopmental-craniofacial syndrome with variable renal and cardiac abnormalities" (Step 1). The user decides to request the addition of a synonym, so they click the "+" button at the right of the Synonyms row, which brings up a form that requests information about the change (Step 2). The user enters the desired new synonym (ZMYM2-related neurodevelopmental disorder with multiple anomalies) and selects the synonym type ("exact") from a pulldown menu. When the form is submitted, BioPortal creates a new GitHub issue in the Mondo repository invoking Ontobot to make the change, and displays a message (third panel) with a hyperlink to let the user go to the issue in GitHub.

When users submit change request proposals, BioPortal collects the data and generates issues that are sent to GitHub in the repository where the ontology source file is maintained. The issues have human readable titles, and the body contents hold a machine-processable string that precisely describes the requested change as a KGCL command, as described in the previous section. **Figure 5** shows an example of how a user can request an ontology change, such as adding a synonym to a term, in BioPortal.

Since a good deal of text across these issues is common, we used Ruby's ERB templating system (24) to build templates for each change request type. For example, a GitHub issue title template for adding a synonym to a class appears in the BioPortal codebase as follows:

```
Proposal: add synonym '<%= synonym_label %>' for <%= concept_label %>
```

After form submission, the template is evaluated and the appropriate fields are dynamically replaced to create the human readable version–for example:

```
Proposal: add synonym 'cortical visual impairment' for cortical blindness
```

BioPortal currently supports four KGCL change request types: synonym creation, synonym removal, class obsoletion, and class renaming. We plan to continue adding support for additional change request types (see **Table 1**). The change request functionality is also configurable on a per-ontology basis, since not all BioPortal users store their ontology source files in GitHub. We currently enable this functionality for four prominent ontologies in BioPortal including Mondo, GO, the Environment Ontology (ENVO) (25), and Uberon.

## Developer tools for working with KGCL

Several tools that we developed or adapted to provide back-end support for Ontobot (including OAK and KGCL-Java) can also be leveraged by advanced users and developers who want to add KGCL support to their applications. Currently not all tools support all features of KGCL.

### Python and OAK

The Ontology Access Kit (OAK) (26) provides both a command line interface and a Python API for two operations: (1) a *diff* operation, which takes as input two ontologies and provides a KGCL diff; (2) an *apply* operation, which takes an ontology plus KGCL commands as input and generates a modified ontology. For both commands, the KGCL CNL and serialized data models in YAML and JSON are supported.

### Java and ROBOT

In parallel to the Python implementation described above, we have also developed a support library for Java, KGCL-Java. The KGCL-Java library provides both a direct translation of the KGCL data model into plain Java classes, which can be used for arbitrary manipulations of KGCL objects in a Java application, and a set of accompanying classes to facilitate working with KGCL. It includes a parser and serializer to convert KGCL objects from and to the KGCL CNL, and classes to implement the changes represented by KGCL objects into OWL axioms so that the changes can be applied to an OWL ontology. Those classes are built on the OWL API library

(27), meaning they can be used to apply KGCL-described changes to any ontology that is in a format supported by the OWL API.

Using the KGCL-Java library, we have also developed a plugin for ROBOT, a command-line tool for automating ontology development tasks (15). The plugin adds a new *apply* command to the ROBOT toolkit. That command allows passing one or more KGCL changes expressed in the KGCL CNL, either directly on the command line or read from a file, and applying them to an ontology as part of a ROBOT pipeline, for example:

```
robot apply -i input.owl -k "obsolete EX:1234 with replacement EX:5678" -o output.owl
```

# Future Work

## Delayed applications of changes

We are considering the possibility of using KGCL to represent provisional changes: changes that are being proposed, but are not yet accepted into the target ontology or KG for some reason (e.g. because the editors need more time to assess whether they are correct). We envision a "provisional mode" for KGCL applications, where the changes that are being described using KGCL syntax are not directly applied to the target ontology, but instead "stored" as KGCL objects within the ontology until they are either approved (in which case they will then be effectively applied) or rejected. The main benefits of such an approach—compared to, for example, keeping provisional changes as PRs waiting to be merged in a repository—is that the changes would be directly visible in the ontology itself, and could be queried and manipulated like any other ontological entity using standard ontology tools and libraries.

How exactly KGCL objects should be stored in an ontology is still under consideration. Our current approach is to represent them as annotations on the very ontological entities that they are intended to modify.

A draft implementation of the "provisional mode" is available in the KGCL-Java library and ROBOT plugin, where a command like `robot apply -k "obsolete EX:1234" --provisional` will store a NodeObsoletion KGCL object into the ontology (instead of actually obsoleting the EX:1234 class), and conversely a command like `robot apply --pending all` will effectively apply all the provisional changes currently stored in the ontology.

## Viewing diffs in BioPortal

KGCL makes it possible to decouple the representation of changes from the process of computing them. It provides a higher-level way to communicate changes that corresponds to how ontologists and curators think of those changes, abstracting away from low-level RDF or OWL diffs. In the future, we plan to use KGCL to support BioPortal's change reporting to show the differences between two versions of an ontology.

## Extension of the core model to support multiple flavors of KGs and ontologies

While our primary use case is changes in ontologies, we have aimed to keep the data model generic enough to use with generalized KGs. We aim to deploy KGCL within our Knowledge Graph Hub (KG-Hub) framework (28), showing at a high level how KGs change over time. For example, the KGCL "edge change" operations are intended to be used with any KG, and are not limited to particular OWL axiom types such as subclass axioms. They are also intended to support property graph style KGs, where individual edges can be annotated with additional contextual information, such as the kind described in the Biolink model (29). However, some users have requested full support for more complex axiom types, including logical definition style equivalence axioms commonly found in OBO ontologies, so we plan to add these in the future.

## AI applications and evaluations

KGCL is fundamentally a tool to help humans communicate about changes in ontologies, either to communicate what changes have been made, or to request desired changes. Generative AI and Large Language Models (LLMs) can complement the use of KGCL for both of these tasks. Our previous work has demonstrated that LLMs can be used under the supervision of expert users to assist with the generation of new terms (30). However, as our community survey showed, many of the bottlenecks in ontology development are around making other kinds of changes to ontologies. To enable the use of LLMs for ontology changes, we developed a prototype ChatGPT plugin that can be used to generate KGCL CNL from free text descriptions of changes (31). This plugin has the limitation that it has to be used within the ChatGPT UI, limiting broader uptake, and making it hard to evaluate. To remedy this, we have started constructing an AI change evaluation set by mining GitHub issues and associated PRs across ontology GitHub repos. For each PR that can be associated with a change, we generate a KGCL CNL description of the changes, together with the issue history associated with that PR. Our intent is to use this in Retrieval Augmented Generation (RAG) applications and to incorporate this into the Ontobot change agent. Our goal is to allow Ontobot to read any issue in a GitHub repo (whether in CNL syntax or plain natural language) and to generate a PR, using the previous history of changes, and information in the ontology as context. The evaluation set and an associated LangChain (32) agent are available from our GitHub repo (33).

## Aligning with related work

Our work is influenced by existing OWL-level diffing tools such as ROBOT and Bubastis. The data model we have devised has similarities to the model used in the COnto-Diff framework, as well as work summarized in Groß et al. (34), the DIACHRON framework (35), and the more recent DynDiff framework (36). The emphasis of our efforts has been on the creation of a human-readable controlled natural language (CNL) that can serve as a means of communication between humans and machines. We have commenced efforts to align and map these data models and to provide bridges between these tools. Simple changes have largely been mapped to their counterparts in COnto-Diff, DIACHRON, and DynDiff. For instance, NodeMove is mapped to: move(c, C_To, C_From), Move_Class(a, B1, B2)/Move_Property(a,

B1, B2), and moveC(c, B1, B2)/moveP(p, B1, B2) accordingly. Specific changes within the KGCL framework, such as NodeMappingChange and PredicateChange, do not have direct equivalents in these models. This is because they focus more on node-related changes (i.e., changes pertaining to classes, properties, and instances) rather than on edge changes or mappings. Instance-level (ABox) changes, like instance addition or deletion, and heuristic changes, such as concept merge and split, have yet to be considered, though as mentioned earlier, the model will be extended to accommodate these types of changes based on community-driven use cases.

## Conclusions

Ontologies and knowledge graphs are highly dynamic in nature, undergoing frequent changes in the light of new knowledge or improved curation. However, change is rarely treated as a first-class object. By providing a standardized representation of changes in ontologies and KGs, KGCL and its associated tooling provide a mechanism to help communicate to curators and users the changes that have occurred in ontologies over time, and a mechanism to communicate and enact desired changes in an ontology.

## Funding

This work was supported by the National Institutes of Health [U24 GM143402, 5U01HG009453-03 (past support)]; the Director, Office of Science, Office of Basic Energy Sciences, of the US Department of Energy [Contract No. DE-AC0205CH11231 to HH, JHC, NLH, CJM], and the National Science Foundation [BBSRC-NSF/BIO BB/T014008/1 to DGG].

## Acknowledgments

In addition to the funding sources mentioned above, a gift from Bosch Corporation also helped support this work.
CJM devised the overall framework and data model and drafted the manuscript
CK and JO devised and wrote the grammar and Python RDF implementation
HH wrote the GitHub adapter and integrated into OAK
NM contributed to the overall framework and data model and helped draft the manuscript
NH contributed substantially to the manuscript
NK contributed to the data model and mappings
JV wrote the BioPortal layer
MM, JG, TR contributed to the overall framework and schema
DGG wrote the Java library and ROBOT adapter
All authors made contributions to the manuscript

We thank all of the participants of the inaugural change language workshop: Allen Baron, Jim Balhoff, Sierra Moxon, Bill Duncan, Sabrina Toro, Matthew Horridge, Nicole Vasilevsky, Naoual Smaili, Emily Hartley, Sue Bello, Yvonne Bradford, Ian Braun, Timothy Redmond, Chris Roeder, Leigh Carmody, Clément Jonquet, Claus Weiland, Jonas Grieb, Thomas Liener, Aleix Puig,

Philip Strömert, Charles Tapley Hoyt, Paul Fabry, Rhiannon Cameron, Damion Dooley, Daniel Olson.

We thank the BioPortal Scientific Advisory Board for their helpful feedback on our project: Yolanda Gil, Clement Jonquet, Andrew Phillips, Andrew Su.## References

1. Diehl AD, Meehan TF, Bradford YM, Brush MH, Dahdul WM, Dougall DS, et al. The Cell Ontology 2016: enhanced content, modularization, and ontology interoperability. J Biomed Semantics. 2016 Jul 4;7(1):44.

2. Mungall CJ, Torniai C, Gkoutos GV, Lewis SE, Haendel MA. Uberon, an integrative multi-species anatomy ontology. Genome Biol. 2012 Jan 31;13(1):R5.

3. Jain S, Pei L, Spraggins JM, Angelo M, Carson JP, Gehlenborg N, et al. Advances and prospects for the Human BioMolecular Atlas Program (HuBMAP). Nat Cell Biol. 2023 Aug;25(8):1089–100.

4. Regev A, Teichmann SA, Lander ES, Amit I, Benoist C, Birney E, et al. The Human Cell Atlas. Elife [Internet]. 2017 Dec 5;6. Available from: http://dx.doi.org/10.7554/eLife.27041

5. Gargano MA, Matentzoglu N, Coleman B, Addo-Lartey EB, Anagnostopoulos AV, Anderton J, et al. The Human Phenotype Ontology in 2024: phenotypes around the world. Nucleic Acids Res. 2024 Jan 5;52(D1):D1333–46.

6. Noy NF, Shah NH, Whetzel PL, Dai B, Dorf M, Griffith N, et al. BioPortal: ontologies and integrated data resources at the click of a mouse. Nucleic Acids Res. 2009 Jul;37(Web Server issue):W170–3.

7. Ong E, Xiang Z, Zhao B, Liu Y, Lin Y, Zheng J, et al. Ontobee: A linked ontology data server to support ontology term dereferencing, linkage, query and integration. Nucleic Acids Res. 2017 Jan 4;45(D1):D347–52.

8. Jupp S, Burdett T, Leroy C, Parkinson HE. A new Ontology Lookup Service at EMBL-EBI. SWAT4LS. 2015;2:118–9.

9. Himmelstein DS, Baranzini SE. Heterogeneous Network Edge Prediction: A Data Integration Approach to Prioritize Disease-Associated Genes. PLoS Comput Biol. 2015 Jul;11(7):e1004259.

10. Callahan TJ, Tripodi IJ, Stefanski AL, Cappelletti L, Taneja SB, Wyrwa JM, et al. An open source knowledge graph ecosystem for the life sciences. Sci Data. 2024 Apr 11;11(1):363.

11. Reese JT, Unni D, Callahan TJ, Cappelletti L, Ravanmehr V, Carbon S, et al. KG-COVID-19: a framework to produce customized knowledge graphs for COVID-19 response. Patterns (N Y). 2020 Aug 18;2(1):100155.

12. Knublauch H, Horridge M, Musen MA, Rector AL, Stevens R, Drummond N, et al. The Protege OWL Experience. In: OWLED [Internet]. 2005. Available from: https://www.researchgate.net/profile/Mark-Musen/publication/221218459_The_Protege_O

WL_Experience/links/09e415113c8d91ab91000000/The-Protege-OWL-Experience.pdf

13. Musen MA, Protégé Team. The Protégé Project: A Look Back and a Look Forward. AI Matters. 2015 Jun;1(4):4–12.

14. Tudorache T, Nyulas C, Noy NF, Musen MA. WebProtégé: A Collaborative Ontology Editor and Knowledge Acquisition Tool for the Web. Semant Web. 2013 Jan 1;4(1):89–99.

15. Jackson RC, Balhoff JP, Douglass E, Harris NL, Mungall CJ, Overton JA. ROBOT: A Tool for Automating Ontology Workflows. BMC Bioinformatics. 2019 Jul 29;20(1):407.

16. Malone J, Stevens R. Measuring the level of activity in community built bio-ontologies. J Biomed Inform. 2013 Feb;46(1):5–14.

17. Binns D, Dimmer E, Huntley R, Barrell D, O'Donovan C, Apweiler R. QuickGO: a web-based tool for Gene Ontology searching. Bioinformatics. 2009 Nov 15;25(22):3045–6.

18. Jacobson M, Sedeño-Cortés AE, Pavlidis P. Monitoring changes in the Gene Ontology and their impact on genomic data analysis. Gigascience [Internet]. 2018 Aug 1;7(8). Available from: http://dx.doi.org/10.1093/gigascience/giy103

19. Hartung M, Groß A, Rahm E. COnto-Diff: generation of complex evolution mappings for life science ontologies. J Biomed Inform. 2013 Feb;46(1):15–32.

20. Moxon S, Solbrig H, Unni D, Jiao D, Bruskiewich R, Balhoff J, et al. The linked data modeling language (LinkML): A general-purpose data modeling framework grounded in machine-readable semantics. In: 2021 International Conference on Biomedical Ontologies, ICBO 2021. CEUR-WS; 2021. p. 148–51.

21. src/kgcl_schema/grammar/kgcl.lark at main · INCATools/kgcl [Internet]. Github; [cited 2024 Sep 17]. Available from: https://github.com/INCATools/kgcl/blob/main/src/kgcl_schema/grammar/kgcl.lark

22. Hoyt CT, Gyori BM. The O3 guidelines: open data, open code, and open infrastructure for sustainable curated scientific resources. Sci Data. 2024 May 29;11(1):547.

23. Decan A, Mens T, Mazrae PR, Golzadeh M. On the Use of GitHub Actions in Software Development Repositories. In: 2022 IEEE International Conference on Software Maintenance and Evolution (ICSME). IEEE; 2022. p. 235–45.

24. erb: An easy to use but powerful templating system for Ruby [Internet]. Github; [cited 2024 Jun 24]. Available from: https://github.com/ruby/erb

25. Buttigieg PL, Pafilis E, Lewis SE, Schildhauer MP, Walls RL, Mungall CJ. The environment ontology in 2016: bridging domains with increased scope, semantic density, and interoperation. J Biomed Semantics. 2016 Sep 23;7(1):57.

26. ontology-access-kit: Ontology Access Kit: A python library and command line application for working with ontologies [Internet]. Github; [cited 2024 Sep 8]. Available from: https://github.com/INCATools/ontology-access-kit

27. Horridge M, Bechhofer S, Noppens O. Igniting the OWL 1.1 touch paper: The OWL API [Internet]. [cited 2024 Sep 9]. Available from: https://ceur-ws.org/Vol-258/paper19.pdf

28. Caufield JH, Putman T, Schaper K, Unni DR, Hegde H, Callahan TJ, et al. KG-Hub-building and exchanging biological knowledge graphs. Bioinformatics [Internet]. 2023 Jul 1;39(7). Available from: http://dx.doi.org/10.1093/bioinformatics/btad418

29. Unni DR, Moxon SAT, Bada M, Brush M, Bruskiewich R, Caufield JH, et al. Biolink Model: A universal schema for knowledge graphs in clinical, biomedical, and translational science. Clin Transl Sci. 2022 Aug;15(8):1848–55.

30. Toro S, Anagnostopoulos AV, Bello S, Blumberg K, Cameron R, Carmody L, et al. Dynamic Retrieval Augmented Generation of Ontologies using Artificial Intelligence (DRAGON-AI) [Internet]. arXiv [cs.AI]. 2023. Available from: http://arxiv.org/abs/2312.10904

31. Creators Mungall C. AI Guided Ontology Curation Workflows and the ROBOT Template GPT helper [Internet]. Available from: https://zenodo.org/records/10901704

32. Topsakal O, Akinci TC. Creating Large Language Model Applications Utilizing LangChain: A Primer on Developing LLM Apps Fast. ICAENS. 2023 Jul 22;1(1):1050–6.

33. Creators Hegde H. LLM Change Agent [Internet]. Available from: https://zenodo.org/doi/10.5281/zenodo.13693477

34. Groß A, Pruski C, Rahm E. Evolution of biomedical ontologies and mappings: Overview of recent approaches. Comput Struct Biotechnol J. 2016 Aug 26;14:333–40.

35. Papavasileiou V, Flouris G, Fundulaki I, Kotzinos D, Christophides V. High-level change detection in RDF(S) KBs. ACM Trans Database Syst. 2013 Apr 26;38(1):1–42.

36. Diaz Benavides S, Cardoso SD, Da Silveira M, Pruski C. Analysis and implementation of the DynDiff tool when comparing versions of ontology. J Biomed Semantics. 2023 Sep 28;14(1):15.